\documentstyle[12pt]{article}
\begin{document}

\baselineskip=8 mm plus 1 mm minus 1 mm

\begin{center}
{\Large \bf The Cosmological Consequences of the Preon Structure of Matter }

\vspace*{.5cm}

{\large  Vladimir Burdyuzha,${}^{*}$  Grigory Vereshkov,${}^{\dag}$} \\
{\large  Olga Lalakulich ${}^{\dag}$ and Yuri Ponomarev ${}^{*}$}

\end{center}

\vspace*{.3cm}

\noindent
${}^{*}$ Astro Space Center of Lebedev Physical Institute of Russian Academy of
Sciences, Profsouznaya str. 84/32, 117810 Moscow, Russia\\
${}^{\dag}$ Rostov State University, Stachki str. 194, 344104 Rostov on Don, Russia

\vspace*{1.5cm}
\begin{abstract}
If the preon structure of quarks, leptons and gauge bosons will be
proved then in the Universe during a relativistic phase transition the
production of nonperturbative preon condensates has occured.
Familons are collective excitations of these condensates. It is shown that
the dark matter consisting of familon type pseudogoldstone bosons
was undergone to two relativistic phase
transitions temperatures of which were different. In the result of these
phase transitions the structurization of dark matter and therefore the
baryon subsystem has taken place. In the Universe two characteristic scales
which have printed this phenomenon arise naturally.
\end{abstract}

\vspace*{1cm}

  The commonly accepted point of view on formation of Large Scale Structure                                                                                                                           (LSS) of the Universe is based on the assumption about de
velopment after
(LSS) of the Universe is based on the assumption about development after
inflation period of homogeneous and isotropic Gausian scalar perturbations
(perturbations of density). The post-recombination spectrum of these
perturbations connects with initial perturbations by the transition function
which depends on nature of dark matter very strong. The standard model
predicts $\Omega_{total} = 1$ that is $\Omega_{CDM} + \Omega_{HDM} +
\Omega_{baryon} + \Omega_{\Lambda} = 1$ in  which large-mass objects are
created at $z \sim  2-3$. For formation of LSS significantly earlier
($z \sim 6-8$) it is necessary to include in $\Omega_{total}$ the density of
vacuum energy ($\Lambda$-term). Our hypothesis about the formation of seeds
for LSS is connected with the late relativistic phase transitions (RPT)
in the gas of familon type pseudo-Goldstone bosons (pseudo-GB) which are a
probable candidate in dark matter (DM).

   The foundation of RPT theory has been formulated by Kirzhnits and Linde and
cosmological application of this theory is used many authors. The time of
the beginning of these RPT depends on a mass of familons. If
$m_{GB} > 3 \times 10^{-4}\ eV$ then the essential influence of these RPT
may take place in no far last evolution of our Universe. If
$m_{GB} < 3 \times 10^{-4}\ eV$ then similar influence of these RPT may take
place in the nearest cosmological future when $T_{CMB} < 2.7\ K$.
We suppose that our particles theory of RPT in the gas and the vacuum of
pseudo-GB allows more effectively to obtain $\delta \rho/\rho \sim 1$ on
earlier stage of the Universe evolution ($z > 5$) than the standard model.

    We have investigated the cosmological consequenses of the simplest
boson-fermion model of quarks and leptons [1]. Our interest to the
preon model was induced by the fact of possible leptoquarks resonance in
the experiment HERA [2]. We have researched more detail the structure
of preon nonperturbative vacuum arising in the result of the
correlation of nonabelian fields on two scales ($\Lambda_{mc} \gg 1
\; Tev$ is the confinement scale of metacolour and $\Lambda_{c} \sim
150 \; Mev$ is QCD scale). We have detected that in spectrum of
excitations of heterogenic nonperturbative preon vacuum
pseudo-GB modes of familon type occurs.  Familons are created in the
result of spontaneous breaking symmetry of quark-lepton generations.
Nonrezo masses of familons are the result of superweak interactions
with quark condensates. We consider these particles as the basic
constituent of cosmological dark matter. The distinguishing
characteristic of these particles is the availability of the residual
$U(1)$ symmetry and possibility of it spontaneous breaking for
temperatures $\frac {\Lambda_{c}}{\Lambda_{mc}} \sim 10^{-3} \; eV$
in the result of RPT.

We have proposed that this relativistic phase transition  has
direct relation to the production of primordial perturbations in DM
the evolution of which leads to baryon large scale structure
formation. The idea of RPT in the cosmological gas of
pseudo-GB in connection with LSS problem was formulated
by Frieman et al. [3]. Here we have investigated by quantitatively
the preon-familon model of this RPT.

At the beginning we point more detail the astrophysical motivation of our
theory. Observational data show that some baryon objects such as the
quasars on $z \sim 4.69$ and $z \sim 4.41$ [4] and  galaxies on $z > 5$ [4]
were produced as minimum on redshifts $z \sim 6 \div 8$. This is the
difficulty for standard CDM and CHDM model to produce their (the best fit is
$z \sim 2 \div 3$ and observations provide the support of this).
If early baryon cosmological structures produced on $z > 10$ then the key role
must play DM particles with nonstandard properties. In standard model
DM consists of ideal gas particles with $m \approx 0$ practically
noninteracting with usual matter (till now they  do not detected because of
their superweak interaction with baryons and leptons).

Certainly a characteristic moment of the most of
cosmological structures formation finishing remains the same
($z \sim 2 \div 3$) and the
appearence of baryon structures on high $z \; (z > 4)$ will be a result of
statistical outburst evolution of the spectrum of DM density
perturbations. Early cosmological baryon structures are connected to
statistical outbursts in sharp nonlinear physical system which is RPT
(the production of inhomogeneities). For clearity we note again that the
appearence of early baryon cosmological structures is the consequense
of evolution of inhomogeneities on boundary of phase domains then as
general situation of RPT in familon gas leads to "natural" picture of the
formation of baryon LSS ($z \sim 2 \div 3$). In frame of our theory some
unclear questions of the astrophysical interpretation of globular clusters
age, distribution of early baryon structures (quasars, CO clouds, galaxies)
may be understood and next generation of space instruments (Next Generation
Space Telescope, Far-Infrared and Submillimeter Space Telescope) will
help to do this.

A new theory of DM must combain properties of  superweak interaction
of DM particles with baryons and leptons and  intensive interaction of
theses particles each other. Such interactions are provided by
nonlinear properties of DM medium. This is the condition for
realization of RPT.

The familon symmetry is experimentally observed (the different
generations of quarks and leptons participate in gauge interactions the
same way). Breaking of this symmetry gives a mass of particles in
different generations. A hypothesis about spontaneous breaking of
familon symmetry is natural and the origin of Goldstone bosons is
inevitably. The properties of any pseudo-GB as and
pseudo-GB of familon type depends on physical
realization of Goldstone modes. These modes can be arisen  from
fundamental Higgs fields or from collective excitations of a heterogenic
nonperturbative vacuum condensate more complex than quark-gluon one in
QCD. The second possibility can realize the theory in which quarks and
leptons are composite that is  the preon model of elementary
particles. If leptoquarks will be detected then two variants of
explanations may be. If leptoquarks resonanse will be narrow and high
then theses leptoquarks come from GUT or SUSY theories. The low and
wide resonanse can be explained by composite particles only (preons).

Thus, in frame of preon theory DM is interpretated as a system of
familon collective excitations of a heterogenic nonperturbative vacuum.
This system consists of 3 subsystems:

1) familons of up-quark type;

2) familons of down-quark type;

3) familons of lepton type.

On stages of cosmological evolution when $T \ll \Lambda_{mc}$ the heavy
unstable familons are absent. Small masses of familons are the result
of superweak interactions of Goldstone fields with nonperturbative
vacuum condensates and therefore familons acquire status of
pseudo-GB.

The value of these masses is limited by astrophysical and laboratory
magnitudes [5]:

$$m_{astrophysical} \sim 10^{-3} \div 10^{-5} \; eV \eqno(1)$$
$$m_{laboratory} \le 10 \; eV$$

The effect of familons mass production corresponds formally
mathemathically the appearence of mass terms in the Lagrangian of
Goldstone fields. From general considerations one can propose that
mass-terms may arise as with "right" as and with "wrong" signs. The sign
of the mass-terms predetermines the destiny of residual symmetry of
Goldstone fields. In the case of "wrong" sign for low temperatures $T <
T_{c} \sim m_{familons} \sim 0.1 \div  10^{5} \; K$ a Goldstone
condensate produces and the symmetry of the familon gas breaks
spontaneously.

The representation about physical nature of familon  excitations
described above is formalized in a theoretical-field model. As example
we discuss a model only one familon subsystem corresponding to up-
quarks of second and third generations. The chiral-familon group of the
model is $SU_{L}(2) \times SU_{R}(2)$. The familon excitations are
described by eight measure (on number of matrix components)
reducible representation of this group factorized on two irreducible
representations $(F, f_{a}); \; (\psi, \varphi_{a})$ which differ each
other by a sign of space chirality. In this model the interaction of quark
fields with familons occurs. However in all calculations quark
fields are represented in the form of nonperturbative quark
condensates. From QCD and a experiment the connection
between quark and gluon condensates is known:

$$\langle 0 \mid \bar{q} q \mid 0 \rangle \approx \frac{1}{12 m_{q}}
\;\;\;\;\; \langle 0 \mid \frac{\alpha_{s}}{\pi} G^{n}_{\mu \nu} G^{\mu
\nu}_{n} \mid 0 \rangle \approx \frac{3 \Lambda^{4}_{c}}{4 m_{q}} \eqno(2)$$

Here: $q = t, c; \;\; m_{c} \sim 1.5 \; Gev; \;\; m_{t} \sim 175 \; Gev; \;\;
\Lambda_{c} \sim 150 \; Mev$.

The spontaneous breaking of symmetry $SU_{L}(2) \times SU_{R}(2)
\rightarrow U(1)$ is produced by vacuum shifts $\langle \psi \rangle =
v; \; \langle f_{3} \rangle = u$. The numerical values $v, u \sim
\Lambda_{mc}$ are unknown. They must be found by experimentally if
our theory corresponds to reality. Parameters $\; u$ and $v$ together with
the value of condensates (2) define numerical values of basic
magnitudes characteristing the familon subsystem. After
breaking of symmetry  $SU_{L}(2) \times SU_{R}(2) \rightarrow U(1)$
light pseudo-GB fields contain the real pseudoscalar field with
the mass:

$$m^{2}_{\varphi^{'}} = \frac{1}{6(u^{2} + v^{2})} \langle 0 \mid
\frac{\alpha_{s}}{\pi} G^{n}_{\mu \nu} G^{\mu \nu}_{n} \mid 0 \rangle \eqno(3)$$
the complex pseudoscalar field with the mass:

$$m^{2}_{\varphi} = \frac{1}{24 v^{2}} \frac{m_{t}}{m_{c}} \langle 0
\mid \frac{\alpha_{s}}{\pi} G^{n}_{\mu \nu} G^{\mu \nu}_{n} \mid 0
\rangle \eqno(4)$$
and complex scalar field the mass square of which is negative:

$$m^{2}_{f} = - \frac{1}{24 u^{2}} \frac{m_{t}}{m_{c}} \langle 0 \mid
\frac{\alpha_{s}}{\pi} G^{n}_{\mu \nu} G ^{\mu \nu}_{n} \mid 0 \rangle \eqno(5)$$

Complex field with masses (4-5) are the nontrivial representation of
residual symmetry of $U(1)$ group, but the real field (3) is the sole
representation of this group. We propose that cosmological DM consists
of particles with these masses and their analogies from the down-
quark-familon and lepton-familon subsystems.

The negative square of mass of complex scalar field means that for

$$T < T_{c(up)} \sim \mid \bar{m}_{f} \mid \sim
\frac{\Lambda^{2}_{c}}{\Lambda_{mc}} \sqrt{m_{t}/m_{c}} \eqno(6)$$
pseudogoldstone vacuum is unstable that is when $T = T_{c(up)}$ in gas
of pseudogoldstone bosons should be the RPT in a state with
spontaneous breaking $U(1)$ symmetry. Two other familon subsystems can
be studied by the same methods. Therefore DM consisting of
pseudo-GB of familon type is a many component heterogenic
system evolving complex thermodynamical way.

In the phase of breaking symmetry every complex field with masses (4-5)
splits on two real fields with different masses. That is the familon
subsystem of up-quark type consists from five kinds of particles with
different masses. Analogous phenomenon takes place in the down-quark
subsystem. The breaking of residual symmetry is when

$$T_{c(down)} \sim \frac{\Lambda^{2}_{c}}{\Lambda_{mc}}
\sqrt{m_{b}/m_{s}} \eqno(7)$$

In the low symmetric phase this subsystem consists also of five kinds
particles with different masses. In our theory the lepton-familon
subsystem is not undergone RPT therefore it consists of particles and
antiparticles of 3 kinds with 3 different masses.

The relativistic phase transitions in familon subsystems must be
described in frame of temperature quantum field theory. It is important
to underline that sufficiently strong interactions of familons each
other provide the evolution of familon subsystem through state of local
equilibrium type. Our estimates have shown that the transition in
nonthermodynamical regime of evolution occurs on stage after RPT even
if RPT took place for temperature $\sim 10^{-3} \; eV$. The
thermodynamics of familon system may be formulated in approximation of the
self-coordinated field. The methods of the RPT theory which will be used by
us are similar to ones of our article [6]. The  unequilibrium Landau functional of states
$F(T, \eta, m_{A})$ depends on symmetry order parameter $\eta$ and
five effective masses of particles $m_{A}, \; A = 1, 2, 3, 4, 5$

$$F(T, \eta, m_{A}) = - \frac{1}{3} \sum_{A} J_{2} (T, m_{A}) + U(\eta,
m_{A}) \eqno(8)$$

Here $J_{2}$ is the characteristic integrals (similar integrals were
used for the description of RPT in our article [7]). The conditions of
extremum of functional on effective masses give the equation of
connection $m_{a} = m_{a}(\eta, T)$ which formal defines the
typical functional Landau $F(T, \eta)$. The condition of minimum of
this functional on the parameter of order

$$\frac{d^{2}F}{d \eta^{2}} = \frac{\partial^{2}F}{\partial \eta^{2}} +
\sum_{A} \frac{\partial^{2}F}{\partial \eta \partial m_{A}}
(\frac{\partial m_{A}}{\partial \eta}) > 0 \eqno(9)$$
\\
is concordanced with equation of state $\partial F/ \partial \eta =
0$ that allows: a) to establish  the kind of RPT, b) to find
thermodynamical  boundary of stability phases, c) to calculate values
of observed magnitudes (energy density, pressure, thermal capacity,
sound velocity et al.) in each phase.

We have detected that RPT in familon gas is RPT of first kind with wide
region of phases coexistence. Therefore in epoch RPT or more
exactly in region phase coexistence the Universe had block-phase
structure containing domains of different phases. The numerical
modelling of this RPT [8] has shown that average contrast of density in the
block-phase structure is $\delta \epsilon/  \epsilon \sim 1$.

The size of domains and masses of baryon and dark matter inside
domains are defined by distance to horison of events $L_{horiz.}$ at
moment of RPT. As it is seen from (6-7) numerical values of these
magnitudes which are important for LSS theory depend on values of unknown
today parameter of preon confainment $\Lambda_{mc}$.

If inhomogeneties appearing during RPT in familon gas have the relation to
observable scales of LSS (10 Mpc) then $\Lambda_{mc} \sim 10^{5} \; Tev$. More
detail estimates today is premature but it is necessary to note that
suggested theory contains two phase transitions and therefore two
characteristic scales of LSS. Here also it is necessary to underline  that
a catastrophic phenomenon in familon gas could not influence on spectrum
of relic radiation even if $ m_{f} \ge 0.4\;eV$ due to superweak interaction
familons with the usual matter but effects connected with the fragmentation
of DM medium may be superimposed at the spectrum of CMB radiation.

Numerical estimates of inhomogeneities parameters
arising as the result of strong interaction of domains LS
and HS phases in region of their contact show that $\delta \epsilon/ \epsilon
\sim 1$ on scale $L \sim 0.1 L_{horison}$ at the moment of the phase
transition.

The more detail publications can be found in [8-9].

\vspace*{1cm}

\begin{center}
{\bf REFERENCES}
\end{center}

\noindent
1. Terazawa H., Chikashige Y., Akama K., 1977,Phys.Rev. D15,480.\\
Fritzsch H., Mandelbaum G. 1981,Phys.Lett.B 102,319.\\
2. Adloff C. et al., 1997, Z.Phys.C74, 191; Breitweg J., 1997, Z.Phys.C74, 207.\\
3. Frieman J.A., Hill C.T., Watkins R., 1991, Preprint Fermilab Pub-91/324-A;\\
Hill C.T., Schramm D.N., Fry J., 1989, Comments of Nucl. and Particle Phys. 19, 25.\\
4. Omont A. et al., 1996, Nature 382,428; Guilloteau et al., 1997, Astron.and
Astrophys.328,L1; Weymann R.J. et al., 1998, Astrophys.J. 505,L95; Hu E.M.,
Cowie L.L.,McMahon R.G.,1998, Astrophys.J. 502,L99; Afanas'ev V.L.,Dodonov S.N.
1999, Nature (submitted)\\
5. Barnett R.M. et al., 1996,Phys.Rev. D54.\\
6. Vereshkov G., Burdyuzha V., 1995, Intern.J.Modern.Phys. A10,1343.\\
7. Burdyuzha V., Lalakulich O., Ponomarev Yu., Vereshkov G.,
1997,Phys.Rev. D55,7340 R.\\
8. Burdyuzha V., Lalakulich O., Ponomarev Yu., Vereshkov G., 1998, Preprint of
Yukawa Institute for Theoretical Physics (YITP-98-51) and hep-ph/9907531\\
9. Terazawa H. 1999, to be published in the Proceedings of the Intern. Conference
on Modern Developments in Elementary Particle Physics, Cairo. Ed. A.Sabry \\

\end{document}